\documentstyle[12pt]{article}
\bibliography{plain}
\title{\bf 2 and 3-dimensional Hamiltonians with Shape Invariance Symmetry}

\vspace{20mm}

\author{M. A. Jafarizadeh $^{a,b,c}$ \thanks{E-mail:jafarzadeh@ark.tabrizu.ac.ir} ,
H. Panahi-Talemi $^{a,}$
\thanks{E-mail:t-panahi@ark.tabrizu.ac.ir}  and E. Faizi $^{a,}$
\thanks{E-mail:msph@ark.tabrizu.ac.ir} \\
\\ $^a${\small Department of Theoretical Physics and Astrophysics,
Tabriz University, Tabriz 51664, Iran.}
\\ $^b${\small Institute
for Studies in Theoretical Physics and Mathematics, Teheran
19395-1795, Iran.}
\\ $^c${\small Pure and Applied Science
Research Center, Tabriz 51664, Iran.}}

\pagebreak

\begin{document}
\maketitle \vspace{10mm}
\begin{abstract}
Via a special dimensional reduction, that is, Fourier transforming
over one of the coordinates of Casimir operator of $su(2)$ Lie
algebra and 4-oscillator Hamiltonian, we have obtained 2 and 3
dimensional Hamiltonian with shape invariance symmetry. Using this
symmetry we have obtained their eigenspectrum. In the mean time we
show equivalence of shape invariance symmetry and Lie algebraic
symmetry of these Hamiltonians.\\ {\bf Keywords: Lie Algebra,
Exactly Solvable, Shape Invariance, Fourier Transformation }.\\
 {\bf PACs
numbers: 05.45.Ra, 05.45.Jn, 05.45.Tp }
\end{abstract}
\newpage

\section{INTRODUCTION}
Exactly solvable quantum Hamiltonians (ESQH) have always attracted
a lot of interest in theoretical physics and mathematical physics.
Hence, construction of exactly solvable models is of great
interest \cite{Khare,Alhassid,Jaf1,Jaf2,Jaf3}. Familiar solvable
potentials (particularly one-dimensional one) have the property of
shape invariance, where this property has played an important role
in calclulation of their determinant by Heat Kernel method
\cite{Jaf4,Jaf5,Jaf6,Jaf7,Jaf8} . For these potentials,
eigenvalues and eigenvectors can be derived using the well known
methods of supersymmetric quantum mechanics together with shape
invariant factorization. The majority of potentials have also been
shown to possess a Lie algebraic symmetry and hence are also
solvable by group theoretical techniques. Actually one can
establish a connection between ESQH with shape invariance symmetry
and ESQH with Lie algebraic symmetry and can show that they are
indeed equivalent \cite{Balan,Jaf9}. One of the authors has
introduced
 some 2 and 3-dimensional shape invariant Hamiltonians
 \cite{Jaf10,Jaf11,Jaf12}. In theses article they have shown that the
shape invariance symmetry of these models is due to the existence
of some Lie algebraic symmetry. Hence, in this article we
construct new 2 and 3-dimensional EQSH with shape invariance
symmetry, where $su(2)$ and  Heisenberg algebra are responsible
for the existence of shape invariance symmetry in them.

This paper is organized as follows: In section II, using the left
and right invariant vector fields of $su(2)$ Lie algebra we first
construct its Casimir operator. Then via Fourier transformation
over one  of the coordinates we construct 2-dimensional
Hamiltonian $H_{q}(\theta,\psi)$ which possess shape invariance
symmetry. Using this symmetry we obtain its eigenspectrum
analytically. In section III, starting with Hamiltonian of
4-oscillator and Fourier transforming over one of the coordinates,
we obtain 3-dimensional Hamiltonian corresponding to motion of a
charged particle in presence of an electric field. We show that
this 3-dimensional Hamiltonian possess a shape invariance symmetry
and using this symmetry we obtain its eigenspectrum. What is so
important in both models is that both Hamiltonians factorize shape
invariantly into a product of second order differential operators.
These second order operators themselves consist of the product of
first order differential operators. The paper ends with a brief
conclusion.

\section{2-dimensional Hamiltonian obtained from \\$SU(2)$ manifold}
\subsection{Left and Right invariant vector field of $SU(2)$}

Considering the following parametrization of $su(2)$ group
manifold \cite{Eguchi} $$
\Lambda=\exp(i\vec{\sigma}.\vec{n}\psi)=A\left(\begin{array}{cc}
         \exp(i\psi) & 0 \\
         0 & \exp(-i\psi)
         \end{array}
   \right)A^{-1}
$$
\begin{equation}
   =\left(\begin{array}{cc}
         \cos(\psi)-i\cos(\theta)\sin(\psi) & -i\sin(\theta)\sin(\psi)\exp(-i\phi) \\
         -i\sin(\theta)\sin(\psi)\exp(i\phi) & \cos(\psi)+i\cos(\theta)\sin(\psi)
         \end{array}\right),
\end{equation}
where $\sigma_{i}$, i= 1, 2 and 3 are Pauli matrices and $\vec{n}$
is a unit vector defined as: $$
\vec{n}=\sin(\theta)\cos(\phi)\vec{i}+\sin(\theta)\sin(\phi)\vec{j}+\cos(\theta)
\vec{k}, $$ and matrix $A$ corresponds to the coherent state
representation of $su(2)$ defined as \cite{Gilmore} : $$
 A=\left(\begin{array}{cc}
         1 & \tau \\
         0 & 1
         \end{array}\right)
         \left(\begin{array}{cc}
         \exp(\frac{\beta}{2}) & 0 \\
         0 &\exp( \frac{-\beta}{2})
         \end{array}\right)
         \left(\begin{array}{cc}
         1 & 0 \\
         -\tau^{\star} & 1
         \end{array}\right),
$$
 with $\tau=\tan(\frac{\theta}{2})\exp(-i\phi)$ and
$\beta=\ln(1+\tau \tau^{\star})$.

In order to obtain the left and right invariant vector field
$su(2)$ manifold with the above parametrization, it is convenient
first to calculate its left and right invariant one form defined
as $\Lambda^{-1}d\Lambda$ and $d\Lambda\Lambda^{-1}$ respectively
\cite{Isham}.

As an example, let us write left invariant one form $$
\Lambda^{-1}d\Lambda=e^{a}_{\alpha}d\xi^{\alpha}\sigma_{a}, $$
where $e^{a}_{\alpha}$ are 3-beins and
$\xi^{\alpha}=(\theta,\phi,\psi)$ are coordinates of
$su(2)$-manifold. Defining the inverse of 3-bein
$e_{a}^{\alpha}=g^{\alpha\beta}\delta_{ab}e^{b}_{\beta}$ with
$g^{\alpha\beta}$ as inverse of metric $g_{\alpha\beta}$: $$
g_{\alpha,\beta}=\left(\begin{array}{ccc}
         1 & 0& 0 \\
         0 & \sin^{2}(\psi) & 0 \\
         0 & 0 & \sin^{2}(\psi)\sin^{2}(\theta)
         \end{array}\right),
$$
 then the left invariant vector field is defined as:
 $$
L_{a}=e_{a}^{\alpha}\frac{\partial}{\partial\xi^{\alpha}}.
 $$
Using the above prescription we obtain the following expression
for left and right invariant vector field of $su(2)$,
respectively,
\begin{equation}
L_{+}=\frac{i}{2}e^{i\phi}\left[\sin(\theta)\partial_{\psi}+(i+\cos(\theta)\cot(\psi))
\partial_{\theta}+(-\cot(\theta)+i\frac{\cot(\psi)}{\sin(\theta)})\partial_{\phi}\right],
\end{equation}
\begin{equation}
L_{-}=\frac{i}{2}e^{-i\phi}\left[\sin(\theta)\partial_{\psi}+(-i+\cos(\theta)\cot(\psi))
\partial_{\theta}+(-\cot(\theta)-i\frac{\cot(\psi)}{\sin(\theta)})\partial_{\phi}\right],
\end{equation}
\begin{equation}
L_{3}=\frac{i}{2}\left(-\cos(\theta)\partial_{\psi}+\sin(\theta)\cot(\psi)\partial_{\theta}
-\partial_{\phi}\right),
\end{equation}
\begin{equation}
R_{+}=\frac{i}{2}e^{i\phi}\left[\sin(\theta)\partial_{\psi}+(-i+\cos(\theta)\cot(\psi))
\partial_{\theta}+(\cot(\theta)+i\frac{\cot(\psi)}{\sin(\theta)})\partial_{\phi}\right],
\end{equation}
\begin{equation}
R_{-}=\frac{i}{2}e^{-i\phi}\left[\sin(\theta)\partial_{\psi}+(i+\cos(\theta)\cot(\psi))
\partial_{\theta}+(\cot(\theta)-i\frac{\cot(\psi)}{\sin(\theta)})\partial_{\phi}\right],
\end{equation}
\begin{equation}
R_{3}=\frac{i}{2}\left(-\cos(\theta)\partial_{\psi}+\sin(\theta)\cot(\psi)\partial_{\theta}
+\partial_{\phi}\right),
\end{equation}
where $ L_{\pm}=L_{1}\pm iL_{2}$ and $ R_{\pm}=R_{1}\pm iR_{2}$.
It is straightforward to show that the left and right invariant
vector field fulfill the following $su(2)$ Lie algebra:
\begin{equation}
[L_{+},L_{-}]=2L_{3},\;\;\;[L_{3},L_{\pm}]=\pm L_{\pm},
\end{equation}
\begin{equation}
[R_{+},R_{-}]=-2R_{3},\;\;\;[R_{3},R_{\pm}]=\mp R_{\pm},
\end{equation}
also, the left and right invariant generators commute with each
other
 \begin{equation}
[\vec{L},\vec{R}]=0.
 \end{equation}
Considering the Casimir operators of $su(2)$ defined as:
 $$
L^2=\frac{1}{2}(L_{+}L_{-}+L_{-}L_{+})+L_{3}^{2},
 $$
and ignoring the scale 1/4, we obtain
\begin{equation}
L^{2}=-\frac{1}{\sin^{2}(\psi)}\partial_{\psi}\sin^{2}(\psi)\partial_{\psi}
-\frac{1}{\sin^{2}(\psi)}
\left(\frac{1}{\sin(\theta)}\partial_{\theta}\sin(\theta)\partial_{\theta}
+\frac{1}{\sin^{2}(\theta)}\partial^{2}_{\phi}\right).
\end{equation}
In obtaining the above formula we have used the left invariant
generators. It is straightforward to show that we can obtain the
same result with right invariant generators too, that means the
 Casimir operator of left and right operator are the same.

\subsection{$H_{q}(\theta,\psi)$ Hamiltonian}

Here through dimensional reduction we show that the above Casimir
operator reduces to a Hamiltonian of motion of a charged particle
in the presence of electric field. Hence, first we make
one-dimensional reduction ( eliminate the coordinate $\phi$ )
through the usual Fourier transformation defined as
\begin{equation}
 \tilde{f}(q)=\frac{1}{\sqrt{2\pi}}\int_{0}^{2\pi}f(\phi)\exp(-i \phi q) d
 \phi,
\end{equation}
over an arbitrary function $f(\phi)$. Obviously the Casimir
operator (2.11) reduces to the following operator
\begin{equation}
L^{2}_{q}(\theta,\psi)=-\frac{1}{\sin^{2}(\psi)}\partial_{\psi}\sin^{2}(\psi)\partial_{\psi}
-\frac{1}{\sin^{2}(\psi)}
\left(\frac{1}{\sin(\theta)}\partial_{\theta}\sin(\theta)\partial_{\theta}
-\frac{q^{2}}{\sin^{2}(\theta)}\right),
\end{equation}
in the Hilbert space of Fourier transformed wavefunctions. In
general the non- relativistic Hamiltonian of a charged particle
over a 2-dimensional manifold with metric $g_{\mu \nu}$ in the
presence of magneto static field $\vec{B}$ with vector potential
$\vec{A}$ and electro static field $\vec{E}$ with scalar potential
$V$ can be written as \cite{Jaf8, Kleinert}
\begin{equation}
H=-\frac{1}{\sqrt{g}}(\partial_{\mu}-iA_{\mu})
(\sqrt{g}g^{\mu\nu}(\partial_{\nu}-iA_{\nu}))+V,
\end{equation}
where $g$ is the determinant of metric $g_{\mu\nu}$. After
similarity transformation of the Casimir operator (2.13) defined
as:
 $$\tilde{L}^{2}_{q}(\theta,\psi)=\sin^{\frac{1}{2}}(\psi)L^{2}_{q}(\theta,\psi)
 \sin^{-\frac{1}{2}}(\psi), $$
 we have
\begin{equation}
\tilde{L}^{2}_{q}(\theta,\psi)=-\frac{1}{\sin(\psi)}\partial_{\psi}\sin(\psi)\partial_{\psi}
-\frac{1}{\sin^{2}(\psi)}
\left(\frac{1}{\sin(\theta)}\partial_{\theta}\sin(\theta)\partial_{\theta}
-\frac{q^{2}}{\sin^{2}(\theta)}\right)+\frac{1}{4}\cot^{2}(\psi)-\frac{1}{2}.
\end{equation}
Comparing the operator (2.15) with the Hamiltonian (2.14) we
obtain
 $$
g_{\psi\psi}=1,g_{\theta\theta}=\sin^{2}(\psi),g_{\psi\theta}=g_{\theta\psi}=0$$
and
\begin{equation}
A_{\psi}=0,\;\;\;
A_{\theta}=\frac{i}{2}\cot(\theta)=d(\frac{i}{2}\ln(\sin(\theta)).
\end{equation}
It is trivial to see that the vector potential given in (2.16)
corresponds to the pure $u(1)$ gauge field, hence it can be
eliminate through the gauge transform $A\longrightarrow
A_{\mu}+\partial_{\mu}\chi$ with gauge function
$\chi=\frac{i}{2}\ln(\sin(\theta))$. After the above gauge
transformation the $su(2)$-Casimir Hamiltonian reduces to
 \begin{equation}
H_{q}(\theta,\psi)\equiv
e^{-\chi}\tilde{L}^{2}_{q}(\theta,\psi)e^{\chi}
=-\frac{1}{\sin(\psi)}\partial_{\psi}\sin(\psi)\partial_{\psi}
-\frac{1}{\sin^{2}(\psi)}\partial^{2}_{\theta}+\frac{q^{2}
-\frac{1}{4}}{\sin^{2}(\psi)\sin^{2}(\theta)}-\frac{3}{4},
\end{equation}
which can be interpreted as a non-relativistic Hamiltonian of a
point particle over \\ 2-dimensional sphere with metric $$
g_{\mu,\nu}=\left(\begin{array}{cc}
         1 & 0 \\
         0 & \sin^{2}(\psi)
         \end{array}\right)
$$ in the presence of electric field with scalar potential $$
V=\frac{q^{2}-\frac{1}{4}}{\sin^{2}(\psi)\sin^{2}(\theta)}-\frac{3}{4}.
$$ Similarly, the left and right invariant vector fields given in
(2.2)-(2.7) take the following form after the above given
operations, namely, dimensional reduction, similarity
transformation and gauge transformation:
$$\tilde{L}^{\prime}_{+}(q)=\tilde{L}_{+}(q)+g(\theta,\psi,q),\;\;\;\;
\tilde{R}^{\prime}_{+}(q)=\tilde{R}_{+}(q)-g^{\star}(\theta,\psi,q),$$
$$\tilde{L}^{\prime}_{-}(q)=\tilde{L}_{-}(q)-g^{\star}(\theta,\psi,q),\;\;\;\;
\tilde{R}^{\prime}_{-}(q)=\tilde{R}_{-}(q)+g(\theta,\psi,q),$$
$$\tilde{L}^{\prime}_{3}(q)=\tilde{L}_{3}(q),\;\;\;\;
\tilde{R}^{\prime}_{3}(q)=\tilde{R}_{3}(q),$$ where
\begin{equation}
\tilde{L}_{+}(q)
=\frac{i}{2}\left(\sin(\theta)\partial_{\psi}+(i+\cos(\theta)\cot(\psi))\partial_{\theta}
+i(q-1)(-\cot(\theta)+i\frac{\cot(\psi)}{\sin(\theta)})\right)
e^{-\frac{\partial}{\partial q}},
\end{equation}
\begin{equation}
\tilde{L}_{-}(q)
=\frac{i}{2}\left(\sin(\theta)\partial_{\psi}+(-i+\cos(\theta)\cot(\psi))\partial_{\theta}
+i(q+1)(-\cot(\theta)-i\frac{\cot(\psi)}{\sin(\theta)})\right)
e^{\frac{\partial}{\partial q}},
\end{equation}
\begin{equation}
\tilde{L}_{3}(q)=\frac{i}{2}\left(-\cos(\theta)\partial_{\psi}
+\sin(\theta)\cot(\psi)\partial_{\theta}-i q\right),
\end{equation}
\begin{equation}
\tilde{R}_{+}(q)
=\frac{i}{2}\left(\sin(\theta)\partial_{\psi}+(-i+\cos(\theta)\cot(\psi))\partial_{\theta}
+i(q-1)(\cot(\theta)+i\frac{\cot(\psi)}{\sin(\theta)})\right)
e^{-\frac{\partial}{\partial q}},
\end{equation}
\begin{equation}
\tilde{R}_{-}(q)
=\frac{i}{2}\left(\sin(\theta)\partial_{\psi}+(i+\cos(\theta)\cot(\psi))\partial_{\theta}
+i(q+1)(\cot(\theta)-i\frac{\cot(\psi)}{\sin(\theta)})\right)
e^{\frac{\partial}{\partial q}},
\end{equation}
\begin{equation}
\tilde{R}_{3}(q)=\frac{i}{2}\left(-\cos(\theta)\partial_{\psi}
+\sin(\theta)\cot(\psi)\partial_{\theta}+i q\right)
\end{equation}
with $g(\theta,\psi,q)$ is:
 $$ g(\theta,\psi,q)=\frac{1}{4}\left(
\cot(\theta) -i\frac{\cot(\psi)}{\sin(\theta)}\right)
e^{\frac{\partial}{\partial q}}, $$
where * means the usual
complex conjugation. With some
 calculation one can show that the above algebra, that is, the
 commutation relations is unchanged under the above mentioned transformation and the
 Hamiltonian $H_{q}(\theta,\psi)$ can be written in terms of
 generators (2.18)-(2.20) in the following form
  $$
H_{q}(\theta,\psi)=\frac{1}{2}\left(\tilde{L}^{\prime}_{+}(q)\tilde{L}^{\prime}_{-}(q)
+\tilde{L}^{\prime}_{-}(q)\tilde{L}^{\prime}_{+}(q)\right)
+\tilde{L}^{\prime}_{3}(q)^{2} .
 $$
Hence, $H_{q}(\theta,\psi)$ is still Casimir operator $su(2)$ Lie
algebra with generator given in (2.18)-(2.23).

\subsection{Algebraic Solution of $H_{q}(\theta,\psi)$ Hamiltonian}

In order to obtain eigenspectrum of Hamiltonian (2.17) by
algebraic method, first we obtain eigenspectrum of the Casimir
operator (2.11). Since this operator commutes with left and right
invariant generators given in (2.10), therefore, we can obtain
representation of $su(2)$ simply by finding simultaneous
eigenfunctions of the set of commuting operators, $(
R_{3},L_{3},L^{2} )$. Denoting their simultaneous eigenfunction by
$\chi^{l}_{m_{L},m_{R}}(\theta,\psi,\phi)$, we can write
\begin{equation}
L^{2}\chi^{l}_{m_{L},m_{R}}(\theta,\psi,\phi)=l(l+1)\chi^{l}_{m_{L},m_{R}}(\theta,\psi,\phi),
\end{equation}
\begin{equation}
L_{3}\chi^{l}_{m_{L},m_{R}}(\theta,\psi,\phi)=m_{L}\chi^{l}_{m_{L},m_{R}}(\theta,\psi,\phi),
\end{equation}
\begin{equation}
R_{3}\chi^{l}_{m_{L},m_{R}}(\theta,\psi,\phi)=m_{R}\chi^{l}_{m_{L},m_{R}}(\theta,\psi,\phi).
\end{equation}
Now solving the difference of the first order differential
equations (2.25) and (2.26) we deduce that
$\chi^{l}_{m_{L},m_{R}}(\theta,\psi,\phi)$ is proportional to
$e^{-i(m_{R}-m_{L})\phi}$, hence we have
$
\;\;\;\;\chi^{l}_{m_{L},m_{R}}(\theta,\psi,\phi)=e^{-i(m_{R}-m_{L})\phi}
f(\theta,\psi)$, where $ f(\theta,\psi)$ can be determined from
the solution of the sum of the equations (2.25) and (2.26), that
is
 \begin{equation}
-i\cos(\theta)\partial_{\psi}f(\theta,\psi)+i\sin(\theta)\cot(\psi)\partial_{\theta}
f(\theta,\psi)=(m_{L}+m_{R})f(\theta,\psi).
 \end {equation}
 Now considering the
highest weight defined by $m_{L}=-m_{R}=l$. this happenes if the
right hand side of the equation (2.27) vanishes, hence it can be
solved by characteristic method which leads to the following
results: $$
\chi^{l}_{l,-l}(\theta,\psi,\phi)=\exp(2il\phi)f^{max}(\sin(\psi)\sin(\theta))
$$
 where $f^{max}$ is an arbitrary function which can be
determined by solving the first order differential equation:
 $$
R_{+}\chi^{l}_{l,-l}(\theta,\psi,\phi)=0,
\;\;\;\;\;L_{+}\chi^{l}_{l,-l}(\theta,\psi,\phi)=0. $$
 Since the highest weight $\chi^{l}_{l,-l}$ belongs to the kernel
of raising operators $R_{+}$ and $L_{+}$, therefore the sum of the
equations (2.2) and (2.5) leads to $$
 u\frac{d f^{max}(u)}{d u}=2lf^{max}(u),
$$ where $u=\sin(\psi)\sin(\theta)$. Therefore, solving the above
equation we obtain $ f^{max}(u)=u^{2l}$, hence
$\chi^{l}_{l,-l}(\theta,\psi,\phi)$ has the following form
\begin{equation}
\chi^{l}_{l,-l}(\theta,\psi,\phi)=e^{2il\phi}(\sin(\psi)\sin(\theta))^{2l}.
\end{equation}
 The other eigenweights can be obtained through the operation
of the lowering operator $ R_{-}$ and $L_{-}$ over the highest
eigenfunction, that is, we have
\begin{equation}
\chi^{l}_{m_{L},m_{R}}(\theta,\psi,\phi)
=(L_{-})^{l-m_{L}}(R_{-})^{l+m_{R}}(e^{2il\phi}(\sin(\psi)\sin(\theta))^{2l}).
\end{equation}

In order to eliminate the coordinate $\phi$, first we transfer the
function $e^{2il\phi}$ to the left hand side of the operators
$R_{-}$ and $L_{-}$ in (2.29), then we get:
 $$
\chi^{l}_{m_{L},m_{R}}(\theta,\psi,\phi)=e^{i(m_{L}-m_{R})\phi}L_{-}(m_{L}-m_{R}+1)
L_{-}(m_{L}-m_{R}+2)...$$
\begin{equation}
\;\;\;\;\;\;\;\;\;...L_{-}(l-m_{R})R_{-}(l-m_{R}+1)...R_{-}(2l)(\sin(\psi)\sin(\theta))^{2l},
\end{equation}
where the operators $L_{-}(m)$ and $R_{-}(m)$ are defined as:
\begin{equation}
L_{-}(m)=\frac{i}{2}\left(\sin(\theta)\partial_{\psi}
+(-i+\cos(\theta)\cot(\psi))\partial_{\theta} +i
m(-\cot(\theta)-i\frac{\cot(\psi)}{\sin(\theta)})\right),
\end{equation}
\begin{equation}
R_{-}(m)=\frac{i}{2}\left(\sin(\theta)\partial_{\psi}+(i+\cos(\theta)\cot(\psi))\partial_{\theta}
+i m(\cot(\theta)-i\frac{\cot(\psi)}{\sin(\theta)})\right).
\end{equation}
Finally, the Fourier transformation of (2.30) leads to
 $$
\chi^{l}_{q,m}(\theta,\psi)=L_{-}(q+1)L_{-}(q+2)...L_{-}(l+\frac{q-m}{2})
R_{-}(l+\frac{q-m}{2}+1)...$$
\begin{equation}
\;\;\;\;\;\;\;\;\;\;...R_{-}(2l)(\sin(\psi)\sin(\theta))^{2l},
\end{equation}
where $q=m_{L}-m_{R}$ and $m=m_{L}+m_{R}$.

Since the left and right invariant generators  commute with each
other, we can exchange these operators in (2.29) before Fourier
transformation, whereas after Fourier transformation, we can use
only the relation $L_{-}(q)R_{-}(q-1)=R_{-}(q)L_{-}(q-1)$. Since
the Hamiltonian  $ H_{q}(\theta,\psi)$ can be obtained from the
relations (2.15) and (2.17) via similarity transformation and
gauge transformation, we have:
\begin{equation}
H_{q}(\theta,\psi)=\exp(-\xi)L^{2}_{q}(\theta,\psi)\exp(\xi),\;\;\;
L^{2}_{q}(\theta,\psi)\chi^{l}_{q,m}(\theta,\psi)=l(l+1)\chi^{l}_{q,m}(\theta,\psi),
\end{equation}
where $\xi=-\frac{1}{2}\ln(\sin(\psi)\sin(\theta)). $ Hence
eigenfunction of Hamiltonian  $H_{q}(\theta,\psi)$ can be written
as:
\begin{equation}
\tilde{\chi}^{l}_{q,m}(\theta,\psi)=\exp(-\xi)\chi^{l}_{q,m}(\theta,\psi).
\end{equation}

\subsection{Shape Invariance Symmetry of $H_{q}(\theta,\psi)$}

Here in this section we show that the Hamiltonian
$H_{q}(\theta,\psi)$ possess both degeneracy and shape invariance
symmetry \cite{Khare,Alhassid,Jaf1}. As it is shown in section
(II.3), functions $\tilde{\chi}^{l}_{q,m}(\theta,\psi)=
(\sin(\psi)\sin(\theta))^{\frac{1}{2}}\chi^{l}_{q,m}(\theta,\psi)$
are eigenfunctions of Hamiltonian $H_{q}(\theta,\psi)$ with the
corresponding eigenvalue $l(l+1)$. Since $|m_{R}|\leq l$ and
$|m_{L}|\leq l$, therefore, $|q| \leq 2l$ and for a given value of
$q$ the parameter $m$ can take the following values:
\begin{equation}
m= \left\{ \begin{array}{cc}
           0,\pm2,\pm4,...,\pm(2l-|q|) &\;\; for\;\; |q|=even, \\
           \pm1,\pm3,...,\pm(2l-|q|) &\;\;for\;\; |q|=odd.
          \end{array}
          \right.
\end{equation}
Since the eigenvalue of Hamiltonian $H_{q}(\theta,\psi)$ is
independent of $m$, therefore it has $(2l+1-|q|)$ degenerate
states for a given  $l$ or given energy $l(l+1)$. To see the shape
invariance symmetry of Hamiltonian $H_{q}(\theta,\psi)$, first we
consider the Fourier transformed left and right invariant vector
fields:
 $$
\hspace{-5 cm} \tilde{L}_{+}(q)\equiv
L_{+}(q-1)e^{-\frac{\partial}{\partial q}}$$  \begin{equation}
=\frac{i}{2}\left(\sin(\theta)\partial_{\psi}+(i+\cos(\theta)\cot(\psi))\partial_{\theta}
+i(q-1)(-\cot(\theta)+i\frac{\cot(\psi)}{\sin(\theta)})\right)e^{-\frac{\partial}{\partial
q}},
\end{equation}
 $$
\hspace{-5 cm} \tilde{L}_{-}(q)\equiv
L_{-}(q+1)e^{\frac{\partial}{\partial q}}$$
 \begin{equation}
=\frac{i}{2}\left(\sin(\theta)\partial_{\psi}+(-i+\cos(\theta)\cot(\psi))\partial_{\theta}
+i(q+1)(-\cot(\theta)-i\frac{\cot(\psi)}{\sin(\theta)})\right)e^{\frac{\partial}{\partial
q}},
\end{equation}
\begin{equation}
\tilde{L}_{3}(q)\equiv L_{3}(q)
=\frac{i}{2}(-\cos(\theta)\partial_{\psi}+\sin(\theta)\cot(\psi)\partial_{\theta}
-i q )
\end{equation}
and $$\hspace{-5 cm} \tilde{R}_{+}(q)\equiv
R_{+}(q-1)e^{-\frac{\partial}{\partial q}}$$  \begin{equation}
=\frac{i}{2}\left(\sin(\theta)\partial_{\psi}+(-i+\cos(\theta)\cot(\psi))\partial_{\theta}
+i(q-1)(\cot(\theta)+i\frac{\cot(\psi)}{\sin(\theta)})\right)e^{-\frac{\partial}{\partial
q}},
\end{equation}
 $$
\hspace{-5 cm} \tilde{R}_{-}(q)\equiv
R_{-}(q+1)e^{\frac{\partial}{\partial q}}$$
 \begin{equation}
=\frac{i}{2}\left(\sin(\theta)\partial_{\psi}+(i+\cos(\theta)\cot(\psi))\partial_{\theta}
+i(q+1)(\cot(\theta)-i\frac{\cot(\psi)}{\sin(\theta)})\right)e^{\frac{\partial}{\partial
q}},
\end{equation}
\begin{equation}
\tilde{R}_{3}(q)\equiv R_{3}(q)
=\frac{i}{2}(-\cos(\theta)\partial_{\psi}+\sin(\theta)\cot(\psi)\partial_{\theta}
+i q ).
\end{equation}
After some tedious algebraic calculation we can derive the
following relation between the above operators
 \begin{equation}
L_{3}(q\pm1)L_{\pm}(q)-L_{\pm}(q)L_{3}(q)=\pm
L_{\pm}(q)
\end{equation}
\begin{equation} R_{3}(q\pm1) R_{\pm}(q)- R_{\pm}(q) R_{3}(q)= \mp
R_{\pm}(q).
\end{equation}
These relations indicate that Hamiltonian $H_{q}(\theta,\psi)$
possesses shape invariance symmetry. Since by acting the operators
$R_{\pm}(q)$ and $L_{\pm}(q)$ on both sides of eigenvalue
equations:
 $$
L^{2}_{q}(\theta,\psi)\chi^{l}_{q,m}(\theta,\psi)=l(l+1)\chi^{l}_{q,m}(\theta,\psi),$$
$$R_{3}(q)\chi^{l}_{q,m}(\theta,\psi)=\frac{m-q}{2}\chi^{l}_{q,m}(\theta,\psi),$$
$$L_{3}(q)\chi^{l}_{q,m}(\theta,\psi)=\frac{m+q}{2}\chi^{l}_{q,m}(\theta,\psi),
$$ we get,
\begin{equation}
R_{\pm}(q)\chi^{l}_{q,m}(\theta,\psi)=A_{\pm}(q,m)\chi^{l}_{q\pm1,m\mp1}(\theta,\psi),
\end{equation}
\begin{equation}
L_{\pm}(q)\chi^{l}_{q,m}(\theta,\psi)=B_{\pm}(q,m)\chi^{l}_{q\pm1,m\pm1}(\theta,\psi),
\end{equation}
with
 \begin{equation}
A_{\pm}(q,m)=\frac{1}{2}\sqrt{(2l\mp(m-q))(2l\pm(m-q)+2)},
\end{equation}
\begin{equation}
B_{\pm}(q,m)=\frac{1}{2}\sqrt{(2l\mp(m+q))(2l\pm(m+q)+2)}.
\end{equation}
 The above relations imply that the pair of operators $(L_{-},R_{+})$
 $[(L_{+},R_{-})]$ map degenerate eigenstates of Hamiltonian
 $H_{q}(\theta,\psi)$ for a given value of $q$ into each other, that is they
 decrease [increase] the quantum number  $m$ by 2 units as follows:
$$
 L_{-}(q+1)R_{+}(q)\chi^{l}_{q,m}(\theta,\psi)
 =A_{+}(q,m)B_{-}(q+1,m-1)\chi^{l}_{q,m-2}(\theta,\psi),$$
 $$L_{+}(q-1)R_{-}(q)\chi^{l}_{q,m}(\theta,\psi)
  =A_{-}(q,m)B_{+}(q-1,m+1)\chi^{l}_{q,m+2}(\theta,\psi). $$
Now introducing the operator $Y_{+}(q):=L_{+}(q-1)R_{-}(q)$ and
$Y_{-}(q):=L_{-}(q+1)R_{+}(q)$ as the  raising and lowering
operators of degenerates states of Hamiltonian
$H_{q}(\theta,\psi)$, we have the following shape invariance like
symmetry between the degenerate states of Hamiltonian
$H_{q}(\theta,\psi)$:
 $$Y_{-}(q)Y_{+}(q)\chi^{l}_{q,m}(\theta,\psi)
 =E(q,m)\chi^{l}_{q,m}(\theta,\psi)$$
 $$Y_{+}(q)Y_{-}(q)\chi^{l}_{q,m+2}(\theta,\psi)=
 E(q,m)\chi^{l}_{q,m+2}(\theta,\psi)$$
where
 $$ E(q,m)=A_{-}(q,m)A_{+}(q,m+2)B_{-}(q+1,m+1)B_{+}(q-1,m+1).$$
  Thus, for a given value of $q$, we can obtain eigenfunction of Hamiltonian
  $H_{q}(\theta,\psi)$ with eigenvalue $l(l+1)$,
  simply by acting the pairs of operators
  $(L_{-},R_{+})[(L_{+},R_{-})]$ over the highest weight [lowest weight],
  where here we have derived the eigenfunction $\chi^{l}_{q,m}(\theta,\psi)$
  by acting the lowering operator over the highest eigenstate as follows:
\begin{equation}
\chi^{l}_{q,m}(\theta,\psi)=k^{-1}(Y_{-}(q))^{\frac{2l-|q|-m}{2}}
\chi^{l}_{q,(2l-|q|)}(\theta,\psi),
\end{equation}
where
 $$ k=B_{-}(q+1,m+1)B_{-}(q+1,m+3)...$$
 $$\times
B_{-}(q+1,2l-|q|-1)A_{+}(q,m+2) A_{+}(q,m+4)... A_{+}(q,2l-|q|).$$
Using the relation (2.33) we can obtain the highest weight
eigenstates for $q>0$ and $q<0$,
$$\chi^{l}_{q,(2l-|q|)}(\theta,\psi)= \left\{
\begin{array}{cc}
                            L_{-}(q+1)L_{-}(q+2)...\\
                            \times L_{-}(0)R_{-}(1)R_{-}(2)
                            ...R_{-}(2l)
                            (\sin(\theta)\sin(\psi))^{2l}
                            & for\;\, q < 0\\
                            R_{-}(q+1)R_{-}(q+2)...R_{-}(2l)
                            (\sin(\theta)\sin(\psi))^{2l}
                                                        & for\;\; q >
                                                        0.
                              \end{array} \right.
$$

 On the other hand pair operator $(L_{+},R_{+}) [ or (L_{-},R_{-})]$
leave the eigenvalue $m$ and $l$ unchanged while they increase
[decrease] the parameter $q$ by 2 units, that is they map
eigenfunction of Hamiltonian corresponding to the same energy with
different $q$ into each other. That is, they map isospectral
Hamiltonian into each other, which nothing but shape invariance.
In order to show this shape invariance symmetry, we act the
related operators over $\chi^{l}_{q,m}(\theta,\psi)$, we then
obtain:
 $$
L_{+}(q+1)R_{+}(q)\chi^{l}_{q,m}(\theta,\psi)
 =A_{+}(q,m)B_{+}(q+1,m-1)\chi^{l}_{q+2,m}(\theta,\psi)$$
 $$L_{-}(q-1)R_{-}(q)\chi^{l}_{q,m}(\theta,\psi)
  =A_{-}(q,m)B_{-}(q-1,m+1)\chi^{l}_{q-2,m}(\theta,\psi), $$
obviously, the combined action of above operators leave the
eigenvalues $m$ and $l$ unchanged while changing the parameter $q$
by 2-units. Hence we define the operator
$X_{+}(q):=L_{+}(q+1)R_{+}(q)$ and
$X_{-}(q):=L_{-}(q+1)R_{-}(q+2)$ as raising and lowering operators
of parameter $q$. Then the shape invariance symmetry means:
 $$X_{-}(q)X_{+}(q)\chi^{l}_{q,m}(\theta,\psi)
 =N(q,m)\chi^{l}_{q,m}(\theta,\psi)$$
 $$X_{+}(q)X_{-}(q)\chi^{l}_{q+2,m}(\theta,\psi)
 =N(q,m)\chi^{l}_{q+2,m}(\theta,\psi)$$
where
 $$
N(q,m)=A_{+}(q,m)A_{-}(q+2,m)B_{+}(q+1,m-1)B_{-}(q+1,m+1)$$ or $$
N(q,m)= \frac{1}{16}(2l-m-q)(2l+m+q+2)$$ $$ \times
\sqrt{(2l-m+q)(2l-m+q+4)(2l+m-q+2)(2l+m-q-2)}.$$
 For fixed
values of energy $l(l+1)$ and given values of $m$, the parameter
$q$ can take the following values
$$q=(2l-|m|),(2l-|m|-2),...,-(2l-|m|-2),-(2l-|m|). $$ Hence
obtaining the highest eigenstates, by solving the following first
order differential equation $$
X_{+}(2l-|m|)\chi^{l}_{(2l-|m|),m}(\theta,\psi)=0$$ where its
integral leads to
 $$ \chi^{l}_{(2l-|m|),m}(\theta,\psi)= \left\{
\begin{array}{cc}
                            L_{-}(2l-m+1)L_{-}(2l-m+2)...L_{-}(2l)
                            (\sin(\theta)\sin(\psi))^{2l}
                            & for\; m < 0,\\
                            R_{-}(2l-m+1)R_{-}(2l-m+2)...R_{-}(2l)
                            (\sin(\theta)\sin(\psi))^{2l}
                                                        & for\; m >
                                                        0.
                              \end{array} \right.
$$
 Therefore using the shape invariance relation, we can obtain
the eigenstates of Hamiltonian $H_{q}(\theta,\psi)$ by consecutive
action of $q$-lowering operator over $q$-highest weight
eigenstate,
$$\chi^{l}_{q,m}(\theta,\psi)=f^{-1}X_{-}(q)X_{-}(q+2)...X_{-}(2l-|m|-4)
X_{-}(2l-|m|-2)\chi^{l}_{(2l-|m|),m}(\theta,\psi)$$
$$f=A_(q+2,m)A_(q+4,m)...A_{-}(2l-|m|,m)$$ $$\times
B_{-}(q+1,m+1)B_{-}(q+3,m+1)... B_{-}(2l-|m|-1,m+1).$$

\section{3-dimensional Hamiltonian obtained from \\ \hspace{3 cm} 4-Oscillators}

\setcounter{equation}{0}

Here in this section using the $su(2)$-parametrization of previous
section, we obtain a special 3-dimensional Hamiltonian from the
 Hamiltonian of 4-oscillator with the same frequency, where we
obtain its spectrum via the  corresponding spectrum of
4-oscillator Hamiltonian. We show that such a  Hamiltonian
possesses shape invariance symmetry. The Hamiltonian of
4-oscillator with same frequency can be written as: $$
H=-\frac{1}{2}\Sigma_{i=0}^{4}(P_{i}^{2}+\frac{1}{2}\omega^{2}x_{i}^{2}).
$$  Now making the following change of variable:
\begin{equation}
  \begin{array}{c}
          x_{1}=-r\sin(\psi)\sin(\theta)\sin(\phi),\\
           x_{2}=r\sin(\psi)\sin(\theta)\cos(\phi),\\
           x_{3}=r\sin(\psi)\cos(\theta),\\
           x_{4}=r\cos(\psi),
   \end{array}
  \end{equation}
where $\psi,\theta,\phi$ are the same coordinates used in the
 parametrization $su(2)$ manifold, the Hamiltonian takes the
following form $$ \hspace{-3cm} H(r,\theta,\psi,\phi)=
-\frac{1}{2}\left[\frac{1}{r^{3}}\partial_{r}r^{3}\partial_{r}\right.$$
\begin{equation}
\left.+\frac{1}{r}\left(\partial^{2}_{\psi}+2\cot(\psi)\partial_{\psi}
+\frac{1}{\sin^{2}(\psi)}(\partial^{2}_{\theta}+\cot(\theta)\partial_{\theta}
+\frac{1}{\sin^{2}(\theta)}\partial^{2}_{\phi})\right)\right]
+\frac{1}{2}\omega^{2}r^{2}.
\end{equation}
Since angular part of the above Hamiltonian is the same, the one
given in (2.11), therefore, its eigenspectrum can be obtained
straightforwardly through routine separation variable into radial
and angular part which we are not interested in it here in this
work. Actually here we are concerned with special Hamiltonian
which can be obtained from this 4-oscillator Hamiltonian, that is,
those Hamiltonians which possess  shape invariance symmetry.

In order to achieve this, we write the above Hamiltonian in terms
of raising and lowering operators defined in the usual way:
\begin{equation}
H=\omega(a_{1}^{\dag}a_{1}+a^{\dag}_{2}a_{2}+a^{\dag}_{3}a_{3}+a^{\dag}_{4}a_{4}+2),
\end{equation}
where $a_{i}(a_{i}^{\dag})$  are defined as: $$
a_{i}=\sqrt{\frac{\omega}{2}}(x_{i}+\frac{1}{\omega}\frac{d}{dx_{i}}),\;\;
a^{\dag}_{i}=\sqrt{\frac{\omega}{2}}(x_{i}-\frac{1}{\omega}\frac{d}{dx_{i}}).
$$  These operators have the following form in radial coordinate
(3.1)
$$a_{1}(a^{\dag}_{1})=\sqrt{\frac{\omega}{2}}\left[-r\sin(\psi)\sin(\theta)\sin(\phi)
+(-)\frac{1}{\omega}\right(-\sin(\psi)\sin(\theta)\sin(\phi)\partial_{r}$$
$$-\frac{1}{r}\cos(\psi)\sin(\theta)\cos(\phi)\partial_{\psi}
-\frac{1}{r}\frac{\cos(\theta)\sin(\phi)}{\sin(\psi)}\partial_{\theta}
-\frac{1}{r}\frac{\cos(\phi)}{\sin(\psi)\sin(\theta)}\partial_{\phi}\left)\right],$$
$$
a_{2}(a^{\dag}_{2})=\sqrt{\frac{\omega}{2}}\left[r\sin(\psi)\sin(\theta)\cos(\phi)
 +(-)\frac{1}{\omega}\right(\sin(\psi)\sin(\theta)\cos(\phi)\partial_{r}$$
$$ +\frac{1}{r}\cos(\psi)\sin(\theta)\cos(\phi)\partial_{\psi}
+\frac{1}{r}\frac{\cos(\theta)\cos(\phi)}{\sin(\psi)}\partial_{\theta}
-\frac{1}{r}\frac{\sin(\phi)}{\sin(\psi)\sin(\theta)}\partial{\phi}
 \left)\right],$$
$$a_{3}(a^{\dag}_{3})=\sqrt{\frac{\omega}{2}}\left[r\sin(\psi)\cos(\theta)
+(-)\frac{1}{omega}\right(\sin(\psi)\cos(\theta)\partial_{r}$$
$$+\frac{1}{r}\cos(\psi)\cos(\theta)\partial_{\psi}
-\frac{1}{r}\frac{\sin(\theta)}{\sin(\psi)}\partial_{\theta})\left)\right],$$
$$a_{4}(a^{\dag}_{4})=\sqrt{\frac{\omega}{2}}\left[r\cos(\psi)+(-)\frac{1}{\omega}
\left(\cos(\psi)\partial_{r}-\frac{1}{r}\sin{\psi}\partial_{\psi}\right)\right].
$$
 Now let us define the set of new operators $A_{i}(A_{i}^{\dag})$,
   i= 1, 2 in terms of $a_{i} (a_{i}^{\dag})$ :
 $$ A_{1}=\frac{1}{\sqrt{2}}(a_{1}+ia_{2}),\;\;\;
A_{1}^{\dag}=\frac{1}{\sqrt{2}}(a^{\dag}_{1}-ia^{\dag}_{2}),$$
$$A_{2}=\frac{1}{\sqrt{2}}(a_{1}-ia_{2}),\;\;\;
A_{2}^{\dag}=\frac{1}{\sqrt{2}}(a^{\dag}_{1}+ia^{\dag}_{2}), $$
where, these new operators have the following differential form in
radial coordinates:
$$A_{1}=\frac{i}{\sqrt{2}}\sqrt{\frac{\omega}{2}}e^{i\phi}\left[r\sin(\psi)\sin(\theta)\right.$$
\begin{equation}
\left.+\frac{1}{\omega}\left(\sin(\psi)\sin(\theta)\partial_{r}
-\frac{1}{r}\cos(\psi)\sin(\theta)\partial_{\psi}
+\frac{1}{r}\frac{\cos(\theta)}{\sin(\psi)}\partial_{\theta}
+\frac{1}{r}\frac{i}{\sin(\psi)\sin(\theta)}\partial{\phi})\right)\right],
\end{equation}
$$A^{\dag}_{1}=\frac{-i}{\sqrt{2}}\sqrt{\frac{\omega}{2}}e^{-i\phi}
\left[r\sin(\psi)\sin(\theta)\right.$$
\begin{equation}
\left.+\frac{1}{\omega}\left(\sin(\psi)\sin(\theta)\partial_{r}
+\frac{1}{r}\cos(\psi)\sin(\theta)\partial_{\psi}
+\frac{1}{r}\frac{\cos(\theta)}{sin(\psi)}\partial_{\theta}
-\frac{1}{r}\frac{i}{\sin(\psi)\sin(\theta)}\partial{\phi}\right)\right],
\end{equation}
$$A_{2}=-\frac{i}{\sqrt{2}}\sqrt{\frac{\omega}{2}}e^{-i\phi}
\left[r\sin(\psi)\sin(\theta)\right.$$
\begin{equation}
\left.+\frac{1}{\omega}\left(\sin(\psi)\sin(\theta)\partial_{r}
+\frac{1}{r}\cos(\psi)\sin(\theta)\partial_{\psi}
+\frac{1}{r}\frac{\cos(\theta)}{\sin(\psi)}\partial_{\theta}
-\frac{1}{r}\frac{i}{\sin(\psi)\sin(\theta)}\partial{\phi})\right)\right],
\end{equation}
$$A^{\dag}_{2}=\frac{i}{\sqrt{2}}\sqrt{\frac{\omega}{2}}e^{i\phi}
\left[r\sin(\psi)\sin(\theta)\right.$$
\begin{equation}
\left.-\frac{1}{\omega}\left(\sin(\psi)\sin(\theta)\partial_{r}
+\frac{1}{r}\cos(\psi)\sin(\theta)\partial_{\psi}
+\frac{1}{r}\frac{\cos(\theta)}{sin(\psi)}\partial_{\theta}
+\frac{1}{r}\frac{i}{\sin(\psi)\sin(\theta)}\partial{\phi}\right)\right].
\end{equation}
It is also straightforward to show that they have the following
commutator relations:
 $$
[A_{i},A^{\dag}_{j}]=\delta_{ij}\;\;,\;\;\;[A_{i},A_{j}]=[A^{\dag}_{i},A^{\dag}_{j}]=0
,\;\;\;i,j=1,2 \;.$$
 The 4-oscillators Hamiltonian (3.3) can be
written in terms of the new oscillators as follows:
\begin{equation}
H=\omega(A^{\dag}_{1}A_{1}+A^{\dag}_{2}A_{2}+a^{\dag}_{3}a_{3}+a^{\dag}_{4}a_{4}+2).
\end{equation}
Now its eigenspectrum can be obtained  by solving the following
eigenvalue equation
\begin{equation}
 H\Psi_{(n_{1},n_{2},n_{3},n_{4})}(r,\theta,\phi,\psi)=E_{(n_{1},n_{2},n_{3},n_{4})}
 \Psi_{(n_{1},n_{2},n{3},n_{4})}(r,\theta,\phi,\psi),
\end{equation}
by the usual algebraic method. Hence its eigenfunction can be
written as:
\begin{equation}
\Psi_{(n_{1},n_{2},n_{3},n_{4})}(r,\theta,\phi,\psi)=N(A^{\dag}_{1})^{n_{1}}
(A^{\dag}_{2})^{n_{2}}(a^{\dag}_{3})^{n_{3}}(a^{\dag}_{4})^{n_{4}}
\exp(-\frac{\omega}{2}r^{2}),
\end{equation}
with $N=\frac{\omega}{\pi\sqrt{n_{1}!n_{2}!n_{3}!n_{4}!}}$ as the
 normalization constant, and energy
$E_{(n_{1},n_{2},n_{3},n_{4})}=(n_{1}+n_{2}+n_{3}+n_{4}+2)\omega$.
Using the differential representation of the operator, the
wavefunction (3.10) can be written in the following form
 $$\Psi_{(n_{1},n_{2},n_{3},n_{4})}(r,\theta,\phi,\psi)=
 N2^{(1/2)(n_{1}+n_{2})}
e^{i(n_{2}-n_{1})\phi}e^{-(1/2)r^{2}}(r\sin(\psi)\sin(\theta))^{(n_{1}+n_{2})}$$
\begin{equation}
\times {\cal H}_{n_{3}}(r\sin(\psi)\cos(\theta)){\cal
H}_{n_{4}}(r\cos(\psi))
\Sigma_{i=0}^{n_{1}}(-1)^{i}i!\left(\begin{array}{c}
                                     n_{1}\\ i
                                     \end{array} \right  )
                                  \left(\begin{array}{c}
                                 n_{2}\\ i
                             \end{array} \right )
                                     (r\sin(\psi)\sin(\theta))^{2i},
\end{equation}
where $\mathcal{H}$$_{n}$ is the  Hermit polynomial of degree $n$
and $\left(\begin{array}{c}
                                     n\\ r
                                     \end{array}\right
                                     )=\frac{n!}{r!(n-r)!}
$ . Now with the same prescription used in the  previous section,
we can eliminate $\phi$, by Fourier transforming over it. Hence,
 by the Fourier transformation over $\phi$, the  4-oscillator Hamiltonian reduces
to the following Hamiltonian: $$H_{m}(r,\theta,\psi)=
-\frac{1}{2}\left[\frac{1}{r^{3}}\partial_{r}r^{3}\partial_{r}\right.$$
\begin{equation}
\left.+\frac{1}{r^{2}}\left(\partial^{2}_{\psi}+2\cot(\psi)\partial_{\psi}
+\frac{1}{\sin^{2}(\psi)}(\partial^{2}_{\theta}+\cot(\theta)\partial_{\theta}
-\frac{m^{2}}{\sin^{2}(\theta)})\right)\right]
+\frac{1}{2}\omega^{2}r^{2},
\end{equation}
where after similarity transformation through function $r^{1/2}$,
it reduces to
$$\tilde{H}_{m}(r,\theta,\psi)=r^{1/2}H_{m}(r,\theta,\psi)r^{-1/2}
=-\frac{1}{2}\left[\frac{1}{r^{2}}\partial_{r}r^{2}\partial_{r}\right.$$
\begin{equation}
\left.+\frac{1}{r^{2}}\left(\partial^{2}_{\psi}+2\cot(\psi)\partial_{\psi}
+\frac{1}{\sin^{2}(\psi)}(\partial^{2}_{\theta}+\cot(\theta)\partial_{\theta}
-\frac{m^{2}}{\sin^{2}(\theta)}\right)\right]
+\frac{1}{2}\omega^{2}r^{2}+\frac{3}{8r^{2}}.
\end{equation}
On the other hand, the Hamiltonian $H_{m}(r,\theta,\psi)$ given by
 (3.12) can be written in the following form
\begin{equation}
 H_{m}(r,\theta,\psi)=\omega(A^{\dag}_{1}(m+1)A_{1}(m)+A^{\dag}_{2}(m-1)A_{2}(m)+a^{\dag}_{3}a_{3}+a^{\dag}_{4}a_{4}+2),
\end{equation}
with
$$A_{1}(m)=\frac{i}{\sqrt{2}}\sqrt{\frac{\omega}{2}}\left[r\sin(\psi)\sin(\theta)\right.$$
\begin{equation}
 \left.+\frac{1}{\omega}\left(\sin(\psi)\sin(\theta)\partial_{r}
-\frac{1}{r}\cos(\psi)\sin(\theta)\partial_{\psi}
+\frac{1}{r}\frac{\cos(\theta)}{\sin(\psi)}\partial_{\theta}
-\frac{1}{r}\frac{m}{\sin(\psi)\sin(\theta)}\right)\right],
\end{equation}
$$A^{\dag}_{1}(m)=-\frac{i}{\sqrt{2}}\sqrt{\frac{\omega}{2}}
\left[r\sin(\psi)\sin(\theta)\right.$$
\begin{equation}
\left.-\frac{1}{\omega}\left(\sin(\psi)\sin(\theta)\partial_{r}
+\frac{1}{r}\cos(\psi)\sin(\theta)\partial_{\psi}
+\frac{1}{r}\frac{\cos(\theta)}{sin(\psi)}\partial_{\theta}
+\frac{1}{r}\frac{m}{\sin(\psi)\sin(\theta)}\right)\right],
\end{equation}
$$A_{2}(m)=-\frac{i}{\sqrt{2}}\sqrt{\frac{\omega}{2}}
\left[r\sin(\psi)\sin(\theta)\right.$$
\begin{equation}
\left.+\frac{1}{\omega}\left(\sin(\psi)\sin(\theta)\partial_{r}
+\frac{1}{r}\cos(\psi)\sin(\theta)\partial_{\psi}
+\frac{1}{r}\frac{\cos(\theta)}{\sin(\psi)}\partial_{\theta}
+\frac{1}{r}\frac{m}{\sin(\psi)\sin(\theta)}\right)\right],
\end{equation}
$$A^{\dag}_{2}(m)=\frac{i}{\sqrt{2}}\sqrt{\frac{\omega}{2}}
\left[r\sin(\psi)\sin(\theta)-\frac{1}{\omega}\right.$$
\begin{equation}
\left.\left(\sin(\psi)\sin(\theta)\partial_{r}
+\frac{1}{r}\cos(\psi)\sin(\theta)\partial_{\psi}
+\frac{1}{r}\frac{\cos(\theta)}{sin(\psi)}\partial_{\theta}
-\frac{1}{r}\frac{m}{\sin(\psi)\sin(\theta)}\right)\right].
\end{equation}
 It is straightforward to derive the following
relation between Hamiltonian (3.12) and operator
$A_{i}(m)(A_{i}^{\dag}(m))$, i= 1, 2:
\begin{equation}
  \begin{array}{c}
            H(m-1)A_{1}^{\dag}(m)-A_{1}^{\dag}(m)H(m)=\omega A_{1}^{\dag}(m),\\
            H(m+1)A_{2}^{\dag}(m)-A_{2}^{\dag}(m)H(m)=\omega A_{2}^{\dag}(m),\\
            H(m+1)A_{1}(m)-A_{1}(m)H(m)=-\omega A_{1}(m),\\
            H(m-1)A_{2}(m)-A_{2}(m)H(m)=-\omega A_{2}(m),
   \end{array}
  \end{equation}
where $H(m):=H_{m}(r,\theta,\psi)$. The above relations indicate
that Hamiltonian (3.12) possesses shape invariance symmetry. To
see this, we consider the Fourier transformation of eigenvalue
equation (3.9):
\begin{equation}
H(m)\Psi_{(n,m,n_{3},n_{4})}(r,\theta,\psi)=
E_{(n,n_{3},n_{4})}\Psi_{(n,m,n_{3},n_{4})}(r,\theta,\psi),
\end{equation}
where $n= n_{1}+n_{2},\; m=n_{2}-n_{1}$ and
$E_{(n,n_{3},n_{4})}=(n+n_{3}+n_{4}+2) \omega$. Since $n_{1}$ and
$n_{2}$ are positive integers, therefore $n$ is also a positive
integer but $m$ is an integer. For a given value of $m$, the
quantum number $n$ can be either even or odd integer, since,
quantum numbers $n_{1}$ and $n_{2}$ vary by the same amount, so
that $m$ remains constant. Actually for some given value of $m$,
the quantum number $n$ can take the following values
$$n=|m|,|m|+2,|m|+4,...\;.$$ On the other hand, in terms of $n$,
the quantum number $m$ can take the following values
 $$m=-n,-n+2,...,n-2,n \;.$$
 It is
interesting to see that energy of Hamiltonian
$H_{m}(r,\theta,\psi)$ is independent of $m$, hence these
Hamiltonians are isospectral which is due to the existence of
shape invariance symmetry as we show below.

Operating the operator $A_{1}^{\dag}(m)$ on both sides of the
eigenvalue relation (3.20) and using the relations (3.19), we get
$$ H(m-1)(A_{1}^{\dag}(m)\Psi_{n,m}(r,\theta,\psi)) =(E_{n}+
\omega)(A_{1}^{\dag}(m)\Psi_{n,m}(r,\theta,\psi)),$$ therefore,
$A_{1}^{\dag}(m)\Psi_{n,m}(r,\theta,\psi))$ corresponds to the
eigenfunction of $H(m-1)$ with corresponding eigenvalue $E_{n+1}$,
that is $$A_{1}^{\dag}(m)\Psi_{n,m}(r,\theta,\psi)=
\sqrt{\frac{n-m}{2}+1}\Psi_{n+1,m-1}(r,\theta,\psi),$$
 where
$\Psi_{n,m}(r,\theta,\psi):=\Psi_{(n,m,n_{3},n_{4})}(r,\theta,\psi)$
and $E_{n}:=E_{(n,n_{3},n_{4})}.$
 Similarly, operating
 $A_{2}^{\dag}(m)$ on both sides of (3.20) and using (3.19) we get:
 $$
H(m+1)(A_{2}^{\dag}(m)\Psi_{n,m}(r,\theta,\psi))
=(E_{n}+\omega)(A_{2}^{\dag}(m)\Psi_{n,m}(r,\theta,\psi)),$$ which
leads to $$A_{2}^{\dag}(m)\Psi_{n,m}(r,\theta,\psi)=
\sqrt{\frac{n+m}{2}+1}\Psi_{n+1,m+1}(r,\theta,\psi).$$
 Also by
acting the operators $A_{1}(m)$ and $A_{2}(m)$ on the eigenvalue
relation (3.20) and using the relations (3.19) we obtain
 $$ H(m+1)(A_{1}(m)\Psi_{n,m}(r,\theta,\psi))
=(E_{n}-\omega)(A_{1}(m)\Psi_{n,m}(r,\theta,\psi)),$$ $$
H(m-1)(A_{2}(m)\Psi_{n,m}(r,\theta,\psi))
=(E_{n}-\omega(A_{2}(m)\Psi_{n,m}(r,\theta,\psi)),$$ which imply
the following relations $$A_{1}(m)\Psi_{n,m}(r,\theta,\psi)=
\sqrt{\frac{n-m}{2}}\Psi_{n-1,m+1}(r,\theta,\psi),$$
$$A_{2}(m)\Psi_{n,m}(r,\theta,\psi)=
\sqrt{\frac{n+m}{2}}\Psi_{n-1,m-1}(r,\theta,\psi).$$

From the above relations we conclude that the pair of operators
$(A_{2}(m),A_{1}^{\dag}(m))$ or \\$( A_{2}^{\dag}(m),A_{1}(m))$
acting at eigenfunction $\Psi_{n,m}(r,\theta,\psi)$ of Hamiltonian
$H(m)$,  give eigenfunction of Hamiltonian $H(m\pm2)$ with same
the energy as follows: $$
A_{2}(m-1)A_{1}^{\dag}(m)\Psi_{n,m}(r,\theta,\psi)=
\frac{1}{2}\sqrt{(n+m)(n-m+2)}\Psi_{n,m-2}(r,\theta,\psi),$$ $$
A_{2}^{\dag}(m+1)A_{1}(m)\Psi_{n,m}(r,\theta,\psi)=
\frac{1}{2}\sqrt{(n-m)(n+m+2)}\Psi_{n,m-2}(r,\theta,\psi).$$ Now
introducing the operators $ A_{-}(m):=A_{2}(m-1)A_{1}^{\dag}(m)$
and $A_{+}(m):=A_{2}^{\dag}(m-1)A_{1}(m-2)$, we have:
 $$
A_{-}(m)A_{+}(m)\Psi_{n,m-2}(r,\theta,\psi)=
E(n,m)\Psi_{n,m-2}(r,\theta,\psi),$$
$$A_{+}(m)A_{-}(m)\Psi_{n,m}(r,\theta,\psi)=
E(n,m)\Psi_{n,m}(r,\theta,\psi),$$ where
 $$
E(n,m)=\frac{1}{4}(n+m)(n-m+2).$$ The above relations show the
existence of shape invariance symmetry between the Hamiltonian
$H(m)$ and $H(m-2)$ with same given eigenvalue $E_{n}$. Hence we
can obtain the eigenfunction $\Psi_{n,m}(r,\theta,\psi)$ of
Hamiltonian $H(m)$ by consecutive action of related raising
operators over $\Psi_{n,n}(r,\theta,\psi)$:
 $$ \Psi_{n,m}(r,\theta,\psi)=c^{-1}
A_{-}(m+2)A_{-}(m+4)...A_{-}(n-2)A_{-}(n)
\Psi_{n,n}(r,\theta,\psi),$$ where
 $$
c=\frac{1}{2^{\frac{n-m}{2}}}\sqrt{(n-m)!!
2n(2n-2)...(n+m+4)(n+m+2)},$$
 and
  $$(n-m)!!=(n-m)(n-m-2)...4\times2,$$
 $$\Psi_{n,n}(r,\theta,\psi)\equiv
\Psi_{(n,n,n_{3},n_{4})}(r,\theta,\psi)=
(a_{3}^{\dag})^{n_{3}}(a_{4}^{\dag})^{n_{4}}A_{2}^{\dag}(n-1)
A_{2}^{\dag}(n-2)...A_{2}^{\dag}(1)A_{2}^{\dag}(0)e^{-\frac{1}{2}\omega
r^{2}}.$$
 Of course we can obtain the eigenfunction
$\Psi_{(n,m,n_{3},n_{4})}(r,\theta,\psi)$ by reduction of
coordinate $\phi$ via Fourier transformation of (3.11), which has
the following form:
 $$\Psi_{(n,m,n_{3},n_{4})}(r,\theta,\psi)=
 N 2^{\frac{n}{2}}e^{-(1/2)r^{2}}(r\sin(\psi)\sin(\theta))^{n}$$
$$ \times {\cal H}_{n_{3}}(\sin(\psi)\sin(\theta)){\cal
H}_{n_{4}}(r\cos(\psi)) \Sigma_{i=0}^{\frac{n-m}{2}}(-1)^{i}
i!\left(\begin{array}{c}
                                    \frac{n-m}{2}\\ i
                                     \end{array}   \right )
                                     \left(\begin{array}{c}
                                     \frac{n+m}{2}\\ i
                                     \end{array}  \right )
                                     (r\sin(\psi)\sin(\theta))^{2i}.
$$

\section{CONCLUSION}
Here in this work having Fourier transformed 3 and 4-dimensional
Hamiltonians associated with $su(2)$ and Heisenberg Lie algebra we
have been able to obtain 2 and 3-dimensional Hamiltonian whit
shape invariance symmetry. It would be interesting to obtain
many-body Hamiltonian in one dimension or higher, which possesses
shape invariance symmetry by appropriate Fourier transformation
over some coordinates of Hamiltonian associated with higher ranks
semisimple and non semisimple Lie algebra. This is under
investigation.

\begin{center}
{\bf {  ACKNOWLEDGEMENT}}
\end{center}
We wish to thank  Dr. S.K.A. Seyed Yagoobi for carefully reading
the article and for his constructive comments.

\end{document}